\documentclass[article]{aa} 
\usepackage{graphicx}
\usepackage{natbib}
\bibpunct{(}{)}{,}{a}{}{;}
\usepackage{amssymb}
\usepackage{amsmath}
\usepackage{url}
\usepackage{units}
\usepackage{multirow,color}
\usepackage{txfonts}
\def\lsf{\lambda_{\rm SF}}
\def\lo{\lambda_0}
\def\li{\lambda_{\rm I}}
\def\lf{\lambda_{\rm flat}}
\def\lii{\lambda_{\rm II}}

\begin{document}
\title{Protostellar half-life: new methodology and estimates}


\author{L.E. Kristensen\inst{1}
\and M.M. Dunham\inst{2, 3}
}

\institute{
Centre for Star and Planet Formation, Niels Bohr Institute and Natural History Museum of Denmark, University of Copenhagen, {\O}ster Voldgade 5-7, DK-1350 Copenhagen K, Denmark, \email{lars.kristensen@nbi.ku.dk}  \and
Department of Physics, State University of New York at Fredonia, 280 Central Ave, Fredonia, NY 14063, USA \and
Harvard-Smithsonian Center for Astrophysics, 60 Garden Street, Cambridge, MA 02138, USA
}

\date{Submitted: \today; Accepted: \today}


\abstract
{Protostellar systems evolve from prestellar cores, through the deeply embedded stage and then disk-dominated stage, before they end up on the main sequence. Knowing how much time protostellar systems spend in each stage is crucial for understanding how stars and associated planetary systems form, because a key constraint is the time available to form such systems. Equally important is understanding what the spread or uncertainty in these inferred time scales is. The most commonly used method for inferring protostellar ages is to assume the lifetime of one evolutionary stage, and then scale this lifetime to the relative number of protostars in the other stages, i.e., the method assumes populations are in steady state. The number-counting method does not take into account the underlying age distribution and apparent stochasticity of star formation, nor that star formation is sequential, i.e., populations are not in steady state. To overcome this, we propose a new scheme where the lifetime of each protostellar stage follows a distribution based on the formalism of sequential nuclear decay. In this formalism, the main assumptions are: Class 0 sources follow a straight path to Class III sources, the age distribution follows a binomial distribution, and the star-formation rate is constant throughout. The results are that the half-life of Class 0, Class I, and Flat sources are (2.4$\pm$0.2)\%, (4.4$\pm$0.3)\%, and (4.3$\pm$0.4)\% of the Class II half-life, respectively, which translates to 47$\pm$4, 88$\pm$7, and 87$\pm$8 kyr, respectively, for a Class II half-life of 2 Myr for protostars in the Gould Belt clouds with more than 100 protostars. The mean age of these clouds is 1.2$\pm$0.1 Myr, and the total inferred star formation rate is (8.3$\pm$0.5)$\times$10$^{-4}$ $M_\odot$ yr$^{-1}$ for a mean protostellar mass of 0.5 $M_\odot$. The critical parameters in arriving at these numbers are the assumed half-life of the Class II stage, and the assumption that the star-formation rate and half-lives are constant. This method presents a first step in moving from steady-state to non-steady-state solutions of protostellar populations.
}


\maketitle

\section{Introduction}

Protostellar systems evolve from prestellar cores, through the deeply embedded stage and then disk-dominated stage, before they end up on the main sequence. Knowing how long a time protostellar systems spend in each stage is important for understanding how stars and associated planetary systems form \citep[e.g.,][]{hartmann16}, as well as understanding what the spread or uncertainty in these inferred lifetimes is. Any such measurement will naturally constrain current models and simulations of star formation. Furthermore, protostars are not likely to all have a single value for the lifetime of each evolutionary stage, rather the lifetime will be a distribution, and models should reproduce not only the mean or median value, but also the spread. 

That protostars have a range of lifetimes is not surprising. For example, the NGC1333 region in the Perseus molecular cloud is often interpreted to have an age of the order of 1 Myr based on studies of pre-main-sequence stars with disks \citep{aspin03, wilking04, hatchell07}. However, if all Class II sources have a lifetime of 2 Myr, there should be no Class III sources in such a young cloud. Clearly some protostellar systems are evolving faster than others, and protostellar lifetimes must follow a distribution rather than being a single value. Furthermore, the protostellar populations cannot be in a steady-state solution if only a million years has passed since star formation began.

The most commonly used method for inferring protostellar lifetimes is to assume the lifetime of all protostars in a given Class, often Class II, Class III, or combined Class II+III stage \citep[see][for a full discussion]{dunham15}. This lifetime is then scaled to the relative number of protostars in other stages, thereby providing a single number for the lifetime of each stage \citep[e.g.,][]{wilking89, kenyon90, evans09}, but does not take into account that the different clouds used in such studies may have different star-formation rates, nor what the star-forming time of each cloud is, nor that star formation is sequential, i.e., the current population of Class II sources come from an earlier generation of Class 0, I, and Flat sources. Furthermore, this number-counting method does not consider that the lifetime of each stage is going to be a distribution, rather than a single number. Finally, this method suffers from the number of low-mass stars in any cloud being small and several clouds are often grouped together to overcome this limitation. Thus, while providing a good zeroth-order estimate of protostellar lifetimes where the populations are in steady state, there is room for improvement. 

One possible method for moving to non-steady-state solutions and overcoming some of the shortcomings listed above is to use the formalism developed for sequential nuclear decay. In this formalism, nucleus A decays to nucleus B, which decays to nucleus C, etc. until a stable state is reached. Each decay is associated with a decay constant, $\lambda$, or half-life, $t_{1/2}$ = ln(2)/$\lambda$. The probability distribution for the number of decays in a given time interval is naturally Poissonian, since any single nuclear decay is independent of all other decays. Similarly, prestellar cores ``decay'' to Class 0 objects, which decay to Class I, Flat, Class II, Class III sources, which finally decays to the final stable main-sequence stage. \citet{evans09} first introduced the concept of half-life as applied to protostars to emphasize that protostars are not born with a single pre-determined lifespan, but that lifetimes form some distribution with an associated half-life. 

To avoid confusion, the following nomenclature will be adopted: age refers to the time a protostar has been in existence, lifetime refers to the time a protostar spends in a given Class. Half-life refers to the timescale over which half the protostars in a Class have a shorter (or longer) lifetime. Finally, the star-forming time is the time since the first star formed in a cloud.

In this paper, we further develop the concept of half-life (distribution, non steady state) as opposed to lifetime (single value, steady state) by setting up the equations for sequential decay and then solving them (Sect. 2). The results and underlying assumptions are discussed in Sect. 3, and the conclusions are presented in Sect. 4.

\section{Methods and results}
\label{sec:obs}

\begin{table}
\caption{Number of protostars in various clouds \citep{dunham15}. \label{tab:obs} }
\begin{center}
\begin{tabular}{l c c c c c c c}\hline\hline 
Cloud 		& 0& I & 0+I	& Flat & II & III & Total \\ \hline
Perseus		& 27	& 39	& 	76		& 35 & 235 & 31\phantom{$^a$} & 377 \\
Serpens		& 9 	& 25	& 	34 		& 18 & 131 & 39\phantom{$^a$} & 222 \\
Aquila		&&& 83 		& 65 & 327 & 63\tablefootmark{a} & 538 \\
Auriga/CMC	&&& 35 		& \phantom{1}8 & \phantom{1}73 & 15\phantom{$^a$} & 131 \\
Cepheus		&&& 18 		& 11 & \phantom{1}61 & 12\phantom{$^a$} & 102 \\
IC5146		&&& 28 		& 10 & \phantom{1}79 & 14\phantom{$^a$} & 131 \\
Ophiuchus	& 3	& 25	& 	28 		& 43 & 174 & 32\phantom{$^a$} & 277 \\ \hline
Total			&&& 326	& 210	& 1245 & 	207 & 	1988 \\
\hline\hline
\end{tabular}
\tablefoot{
\tablefoottext{a}{Assuming 90\% of the Class III protostars reported by \citet{dunham15} are AGB contaminants.}
}
\end{center}
\end{table}

In order to estimate protostellar half-lives, a number of assumptions are made:
\begin{enumerate}
\item Evolution (star formation) is a continuous process, i.e., Class 0 objects form continuously in a cloud for as long as the cloud exists, and  Class 0 objects form at some, constant star-formation (core-formation) rate over the star-forming time of the cloud. 
\item Protostellar half-lives only depend on time, not on mass, environment, or other factors. 
\item The evolution of a single protostar is sequential, that is, it will start in Class 0, then go to Class I, Flat, Class II, and finally Class III. 
\item The star-formation rate and half-lives are constant.
\item The observed populations are complete above a certain luminosity limit. 
\item There are no main-sequence stars in a cloud and the final stable state is Class III. 
\end{enumerate}
Most of these assumptions also apply to the number-counting method, and their validity will be discussed below (Sect. 3.1). 

The basic equation to solve is
\begin{align}
{{\rm d} N_0\over {\rm d}t} = \lambda_\mathrm{SF} - \sum_\mathrm{Class\,0s}^{N_0} P_{0-{\rm I}}(t)\ ,
\end{align}
where $P_{0-{\rm I}}(t)$ is the probability of a decay from Class 0 to I at time $t$, the sum is over all Class 0 objects, and $\lsf$ is the formation rate of Class 0 objects. If the probability is assumed to be constant over time and equal for all Class 0 objects, $\lo$, the equation becomes
\begin{equation}
\frac{{\rm d}N_0}{{\rm d}t} = \lsf- \lo N_0
\end{equation}
where $\lo$ is the Class 0 ``decay'' constant and the associated half-life is $t_{1/2}$ = ln(2)/$\lambda$. 

Similarly, the number of Class I, flat, and Class II objects can be written in a form similar to Eq. 1, which becomes the following under the assumption of constant decay rates:
\begin{align}
\frac{{\rm d}N_{\rm I}}{{\rm d}t} &= \lo N_0 - \li N_{\rm I} \\
\frac{{\rm d}N_{\rm flat}}{{\rm d}t} &= \li N_{\rm I} - \lf N_{\rm flat} \\
\frac{{\rm d}N_{\rm II}}{{\rm d}t} &= \lf N_{\rm flat} - \lii N_{\rm II}
\end{align}
where indices I, flat, II refer to Class I, Flat, and Class II sources, respectively. This set of coupled differential equations may be solved analytically and solutions were obtained using the free solver at \url{http://www.wolframalpha.com}. Each of the equations 2--5 is solved with the initial condition $N_i$(0) = 0. The solutions are:
\begin{align}
N_0(t) =& \frac{\lsf}{\lo}\left(1 - e^{- \lo t} \right) \\
N_{\rm I}(t) =& \frac{\lsf}{\li} \left( 1 -  \frac{\li}{\li-\lo} e^{-\lo t} - \frac{\lo}{\lo - \li } e^{-\li t} \right) \\
N_{\rm flat}(t) =& \frac{\lsf}{\lf} \left( 1 - \frac{\li \lf}{\left( \lf - \lo \right) \left( \li - \lo \right)} e^{-\lo t} \right. \nonumber\\
&- \frac{\lo \lf}{\left( \lf - \li \right) \left( \lo - \li \right)} e^{-\li t} \nonumber \\
& \left. - \frac{\lo \li}{\left( \lo - \lf \right) \left( \li - \lf \right)} e^{-\lf t} \right) \\
N_{\rm II}(t) =& \frac{\lsf}{\lii} \left( 1 - \frac{ \li \lf \lii}{\left( \li - \lo \right)\left( \lf - \lo \right)\left( \lii - \lo \right)} e^{-\lo t} \right. \nonumber \\
&- \frac{\lo \lf \lii}{\left( \lo - \li \right)\left( \lf - \li \right)\left( \lii - \li \right)} e^{-\li t} \nonumber \\
&- \frac{\lo \li \lii}{\left( \lo - \lf \right)\left( \li - \lf \right)\left( \lii - \lf \right)} e^{-\lf t} \nonumber \\
& \left. - \frac{\lo \li \lf}{\left( \lo - \lii \right)\left( \li - \lii \right)\left( \lf - \lii \right)} e^{-\lii t} \right) \ .
\end{align}
More generally, the solution to the population in Class $D$ may be written as 
\begin{align}
N_D(t) =& \frac{\lsf}{\lambda_D} \left( 1 - \sum_{i=1}^D c_i e^{-\lambda_i t} \right ) \ ,\\
c_i =& \prod_{j=1, i\ne j}^D \frac{\lambda_j}{\lambda_j - \lambda_i} \ .
\end{align}
The solution is very similar to that found by \citet{bateman10} for consecutive nuclear decay, apart from the constant formation rate of Class 0 objects. We note that if the number of main-sequence stars in a cloud were known, it would be possible to set up an equation for $N_{\rm III}(t)$ and solve it on equal footing with Eq.s 6--9 without assuming that Class III sources are the final stable end product. It is not possible to do so at the moment: solving such an equation for $\lambda_{\rm III}$ would lead to the solution being infinity because the number of stable end products, the main sequence stars, is zero or unknown. 

To further solve these equations for the unknown decay constants, the boundary conditions are:
\begin{align}
N_0(t_0) &= N_0^{\rm obs} \\
N_{\rm I}(t_0) &= N_{\rm I}^{\rm obs} \\
N_{\rm flat}(t_0) &= N_{\rm flat}^{\rm obs} \\
N_{\rm II}(t_0) &= N_{\rm II}^{\rm obs} \\
N_{\rm III}(t_0) &= N_{\rm III}^{\rm obs}
\end{align}
where superscript ``obs'' refers to the observed number of protostars in each Class, and $t_0$ is the unknown star-forming time of the cloud. The total number of protostars in a cloud is thus $N_{\rm tot}(t_0) = \lambda_{\rm SF}\, t_0$.

The number of protostars in each evolutionary Class are adopted from \citet{dunham15}, and they were observed as part of the ``Cores to Disks'' (c2d) \textit{Spitzer} survey \citep{evans09}. To separate Class 0 and I protostars from the common category of 0+I, we follow \citet{dunham15} and assign relative fractions of 35\% and 65\%, respectively \citep{enoch09}. Furthermore, only clouds with many protostars ($N_{\rm tot}$ $>$ 100) are included to obtain the most precise numbers possible for any individual cloud. Finally, the number of Class III objects is unusually high toward Aquila. \citet{dunham15} argues that the reason is an unusually high contamination by AGB stars, since this cloud is located toward the Galactic Center, where the concentration of such stars will be naturally higher. We here adopt a contamination rate of 90\% following \citet{dunham15}. The number of protostars in each cloud and in each Class are summarized in Table \ref{tab:obs}. Finally, the total number of sources in all clouds are considered as a single cloud. 

For each cloud, Equations 6--9 are solved numerically using an iterative bisector method, i.e., the decay rates are not a fit but a solution to the equations. The solutions are degenerate, because we are attempting to constrain six variables ($\lsf$, $\lo$, $\li$, $\lf$, $\lii$, $t_0$) through five observables ($N_0^{\rm obs}$, $N_{\rm I}^{\rm obs}$, $N_{\rm flat}^{\rm obs}$, $N_{\rm II}^{\rm obs}$, $N_{\rm III}^{\rm obs}$). To overcome this obstacle, decay rates are normalized to an assumed Class II half-life of 2 Myr, $\lii$ = ln(2)/2 Myr = 0.347 Myr$^{-1}$. 

To estimate the uncertainty on the half-lives, we make use of assumption 2, that time is the only factor in determining the evolutionary state of a protostar. If that is the case, the protostellar distribution, i.e., number of protostars in a given Class, follows a binomial distribution for each Class \citep[e.g.,][]{huestis02}. We draw 1000 random populations for each cloud, that is, we draw a random number of protostars in each Class where the distribution of these random numbers follows a binomial distribution. This distribution is described by two numbers, the number of draws and the probability. In this case, the number of draws is the total number of sources, $N_{\rm tot}^{\rm obs} = N_0^{\rm obs} + N_{\rm I}^{\rm obs} + N_{\rm flat}^{\rm obs} + N_{\rm II}^{\rm obs} + N_{\rm III}^{\rm obs}$, and the probability is $N_{i}^{\rm obs}$ / $N_{\rm tot}^{\rm obs}$. Here, $N_{i}^{\rm obs}$ is the observed number of protostars in each Class $i$. Next, the modified Bateman equations are solved for each population, providing new values for the decay rates. 1000 populations is a good compromise between computing time and converging on a mean and standard deviation; we tested for convergence by drawing 2000 and 5000 populations for the case of Perseus, and saw no significant variation in either parameter. 

\begin{figure}
\center\includegraphics[width=\columnwidth]{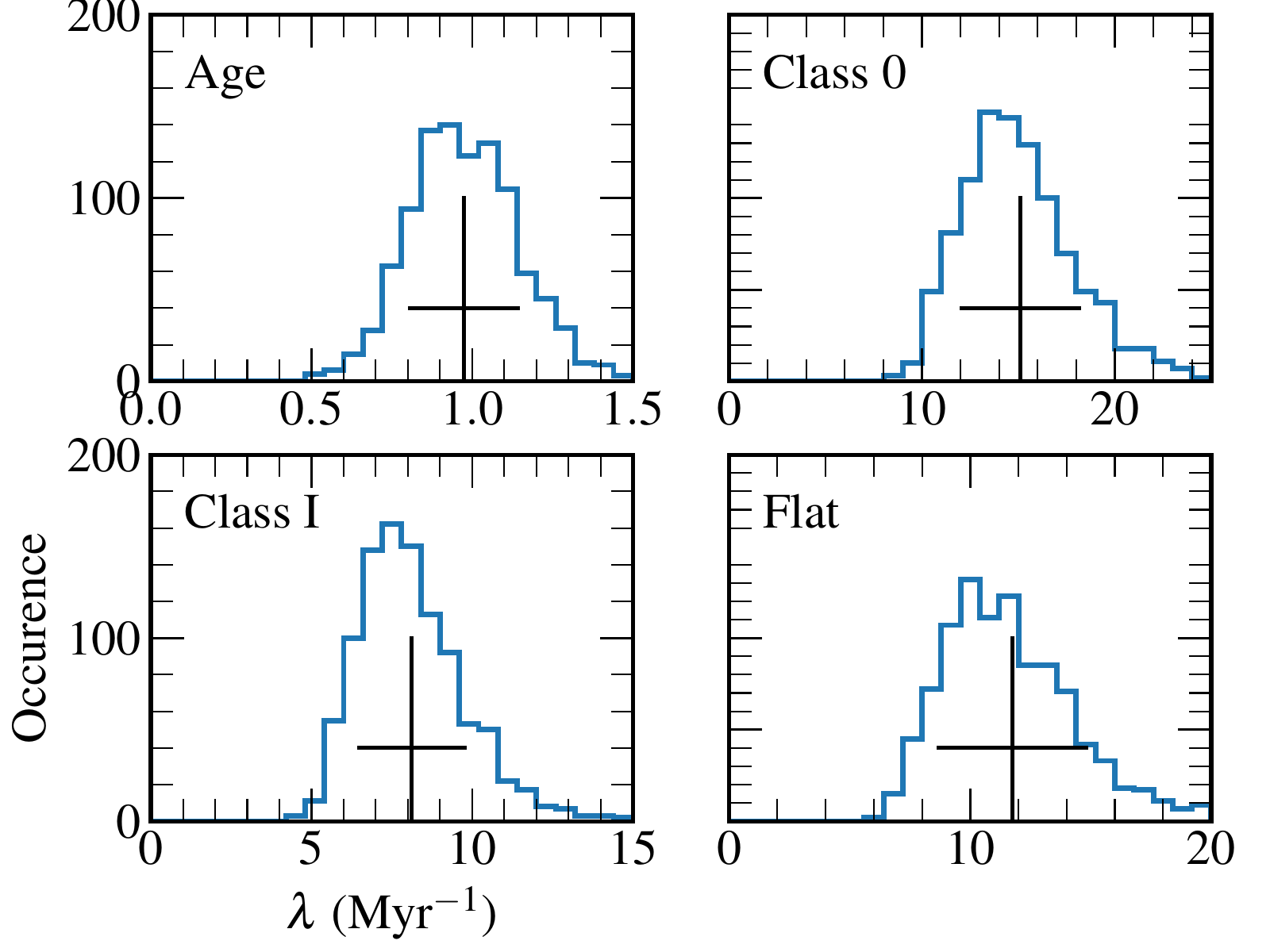}
\caption{Resulting decay rate distributions for 1000 random input distributions for each evolutionary class in Perseus. The vertical lines indicate the mean of each Class decay rate, and the horizontal line is the standard deviation. The x axis in the top left plot is the star-forming time, $t_0$, in units of Myr.  
\label{fig:dist}}
\end{figure}

\begin{figure}
\center\includegraphics[width=0.9\columnwidth]{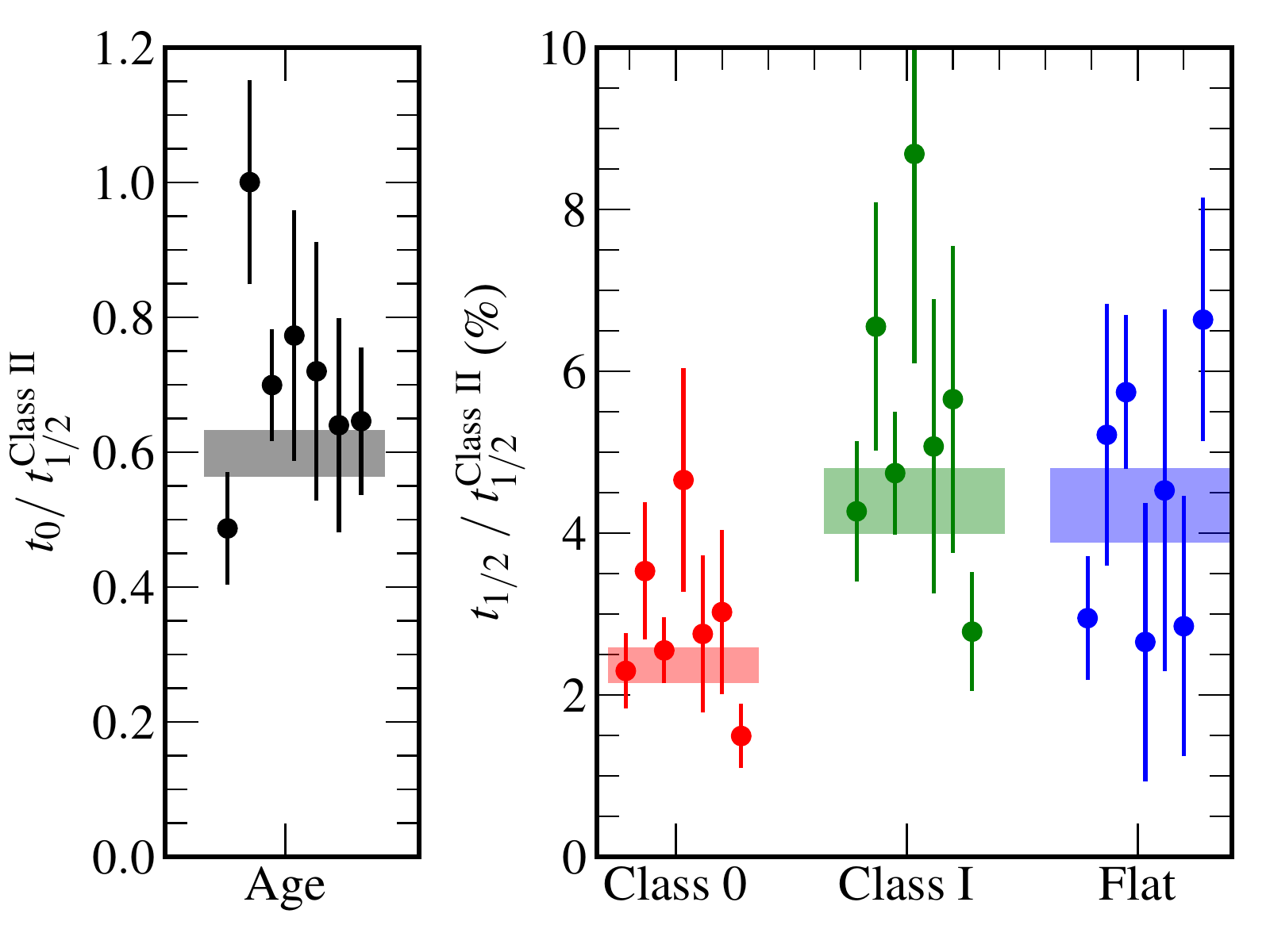}
\caption{Star-forming times (left) and half-lives (right), both relative to the Class II half-life, inferred for protostars in all clouds as a function of evolutionary stage as presented in Tables \ref{tab:age} and \ref{tab:decay}. The clouds have been shifted horizontally for clarity, and are (from left to right): Perseus, Serpens, Aquila, Auriga/CMC, Cepheus, IC5146, and Ophiuchus. The average, when all stars are included, is shown as a shaded box, where the height of the box is the spread. 
\label{fig:lambda}}
\end{figure}

An example is shown in Fig. \ref{fig:dist} for the Perseus Cloud. The resulting distributions are typically slightly skewed with tails extending to higher values of $\lambda$, characteristic of a Poisson distribution. This is not surprising: the binomial distribution tends toward the Poisson distribution for large numbers. The solutions are reported in Table \ref{tab:decay} and displayed in Fig. \ref{fig:lambda}. The inferred half-lives for Class 0, I, and Flat sources all scale linearly with the assumed Class II half-life, so what is important is the ratio of the half-life of a given Class to the half-life of Class II; for this reason, the relative half-lives are also reported in Table \ref{tab:decay}. The star-formation times, $t_0$, are provided in Table \ref{tab:age}. Assuming a mean stellar mass of 0.5 $M_\odot$ corresponding to the peak of the initial mass function \citep[IMF,][]{chabrier03}, the star-formation rate, SFR, may also be inferred as SFR = $N_{\rm tot} \times \left< M \right>  / t_0$. 

The results are that the decay rates are remarkably similar for all clouds, and these agree with the values found for the total set of protostars. The half-lives for Class 0, I, and Flat sources are (2.4$\pm$0.2)\%, (4.4$\pm$0.4)\%, and (4.3$\pm$0.4)\% of the Class II half-life, respectively, which corresponds to 47$\pm$4 kyr, 88$\pm$7 kyr, and 87$\pm$8 kyr, respectively, for a Class II half-life of 2 Myr. The cloud star-formation times are 1--2 Myr, and the total inferred SFR is 830$\pm$50 $M_\odot$ Myr$^{-1}$ for a mean stellar mass of 0.5 $M_\odot$. 

\begin{table}
\caption{Decay rates for each class of protostars. \label{tab:decay} }
\begin{center}
\begin{tabular}{l c c c c }\hline\hline 
 & $\lambda_{0}$ & $\lambda_{\rm I}$	& $\lambda_{\rm Flat}$ \\
Cloud & (Myr$^{-1}$) & (Myr$^{-1}$) & (Myr$^{-1}$) \\ \hline 
Perseus		& 15.1$\pm$3.0 & \phantom{1}8.1$\pm$1.6 & 11.8$\pm$3.1 \\
Serpens		& \phantom{1}9.8$\pm$2.4 & \phantom{1}5.3$\pm$1.2 & \phantom{1}6.6$\pm$2.1 \\
Aquila		& 13.6$\pm$2.2 & \phantom{1}7.3$\pm$1.2 & \phantom{1}6.0$\pm$1.0 \\
Auriga/CMC	& \phantom{1}7.4$\pm$2.2 & \phantom{1}4.0$\pm$1.2 & 13.1$\pm$8.5 \\
Cepheus		& 12.6$\pm$4.4 & \phantom{1}6.8$\pm$2.5 & \phantom{1}7.7$\pm$3.8 \\
IC5146		& 11.5$\pm$3.8 & \phantom{1}6.1$\pm$2.0 & 12.2$\pm$6.9 \\
Ophiuchus	& 23.2$\pm$6.2 & 12.5$\pm$3.3 & \phantom{1}5.2$\pm$1.2 \\ \hline 
Total 		& 14.7$\pm$1.1 & \phantom{1}7.9$\pm$0.6 & \phantom{1}8.0$\pm$0.7 \\
$t_{1/2}$ (kyr)	& 47$\pm$4 & 88$\pm$7 & 87$\pm$8 \\
$t_{1/2} / t_{1/2}^{\rm Class\ II}$ (\%)	& \phantom{1}2.4$\pm$0.2 & \phantom{1}4.4$\pm$0.3 & \phantom{1}4.3$\pm$0.4 \\
\hline\hline
\end{tabular}
\end{center}
\end{table}

\begin{table}
\caption{Cloud star-forming times for a Class II half-life of 2 Myr, and star formation rates for a median star mass of 0.5 $M_\odot$. \label{tab:age} }
\begin{center}
\begin{tabular}{l c c}\hline\hline 
Cloud 		& Age (Myr) & SFR ($M_\odot$ Myr$^{-1}$) \\ \hline
Perseus		& 0.97$\pm$0.17 & 194$\pm$33 \\
Serpens		& 2.00$\pm$0.30 & \phantom{1}55$\pm$8\phantom{1} \\
Aquila		& 1.40$\pm$0.17 & 192$\pm$23 \\
Auriga/CMC	& 1.55$\pm$0.37 & \phantom{1}42$\pm$10 \\
Cepheus		& 1.44$\pm$0.38 & \phantom{1}35$\pm$9\phantom{1} \\
IC5146		& 1.28$\pm$0.32 & \phantom{1}51$\pm$13 \\
Ophiuchus	& 1.29$\pm$0.22 & 107$\pm$18 \\ \hline
Total			& 1.20$\pm$0.08 & 831$\pm$53 \\ 
\hline\hline
\end{tabular}
\end{center}
\end{table}

\begin{figure}
\center\includegraphics[width=\columnwidth]{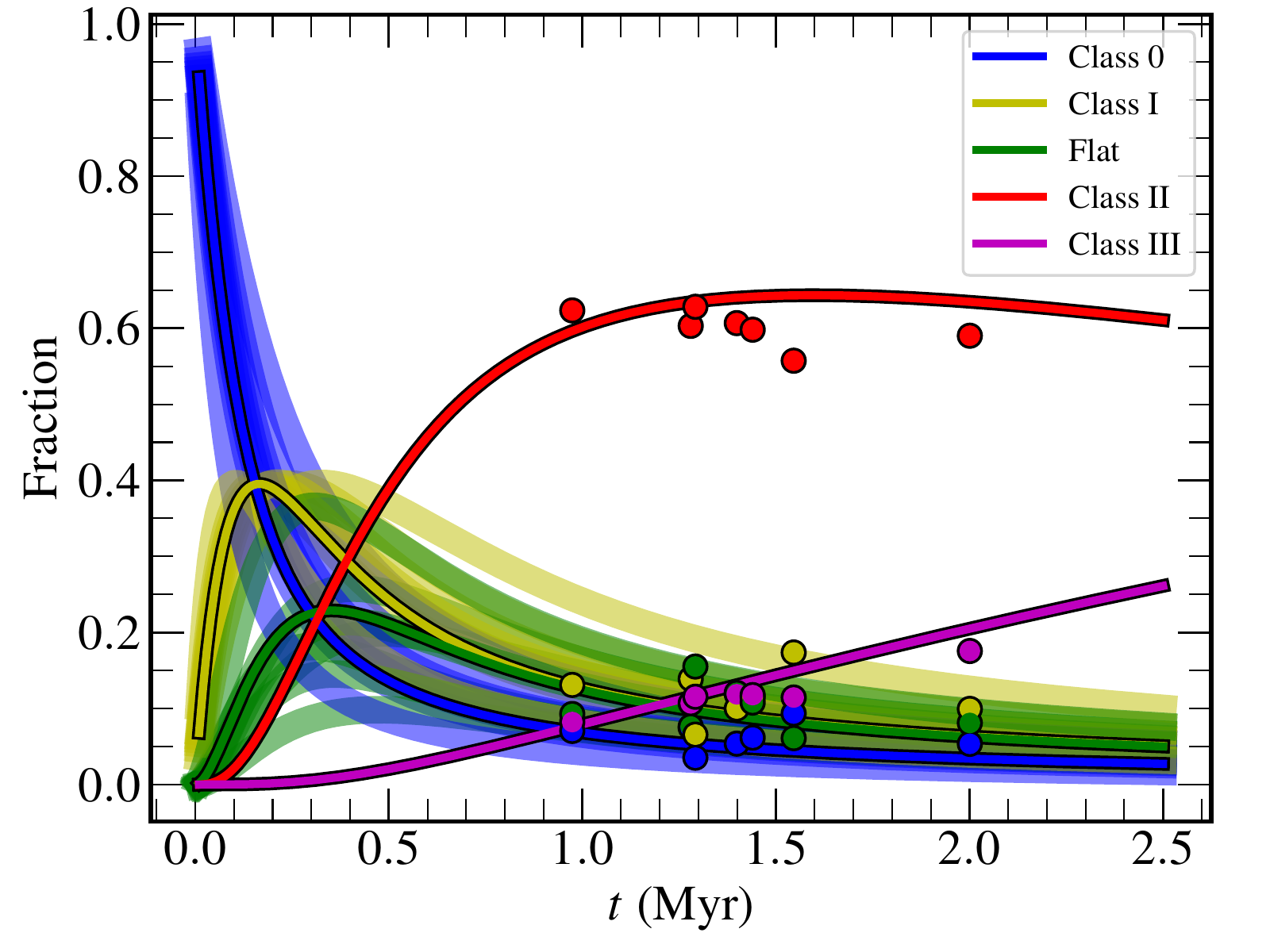}
\center\includegraphics[width=\columnwidth]{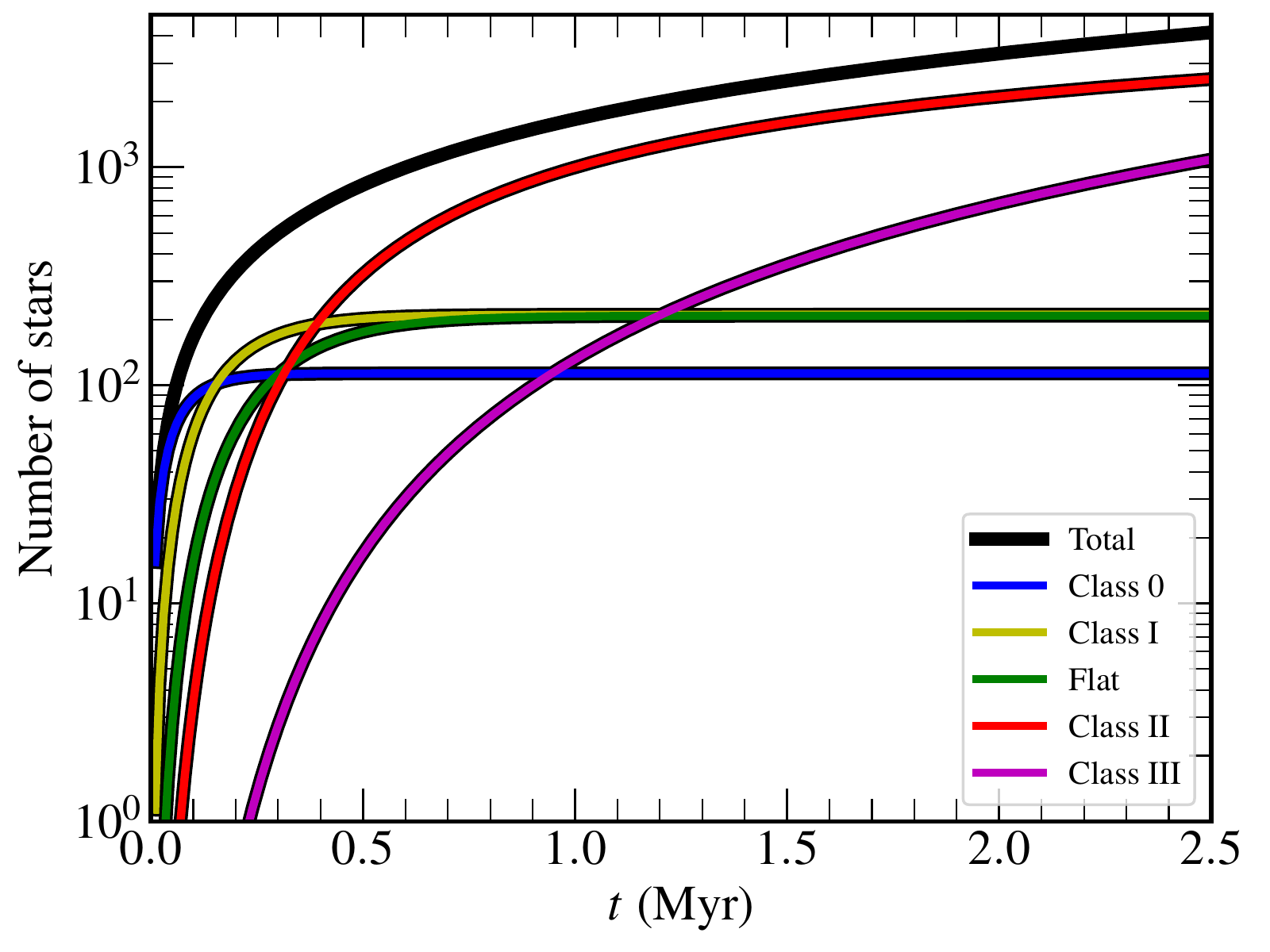}
\center\includegraphics[width=\columnwidth]{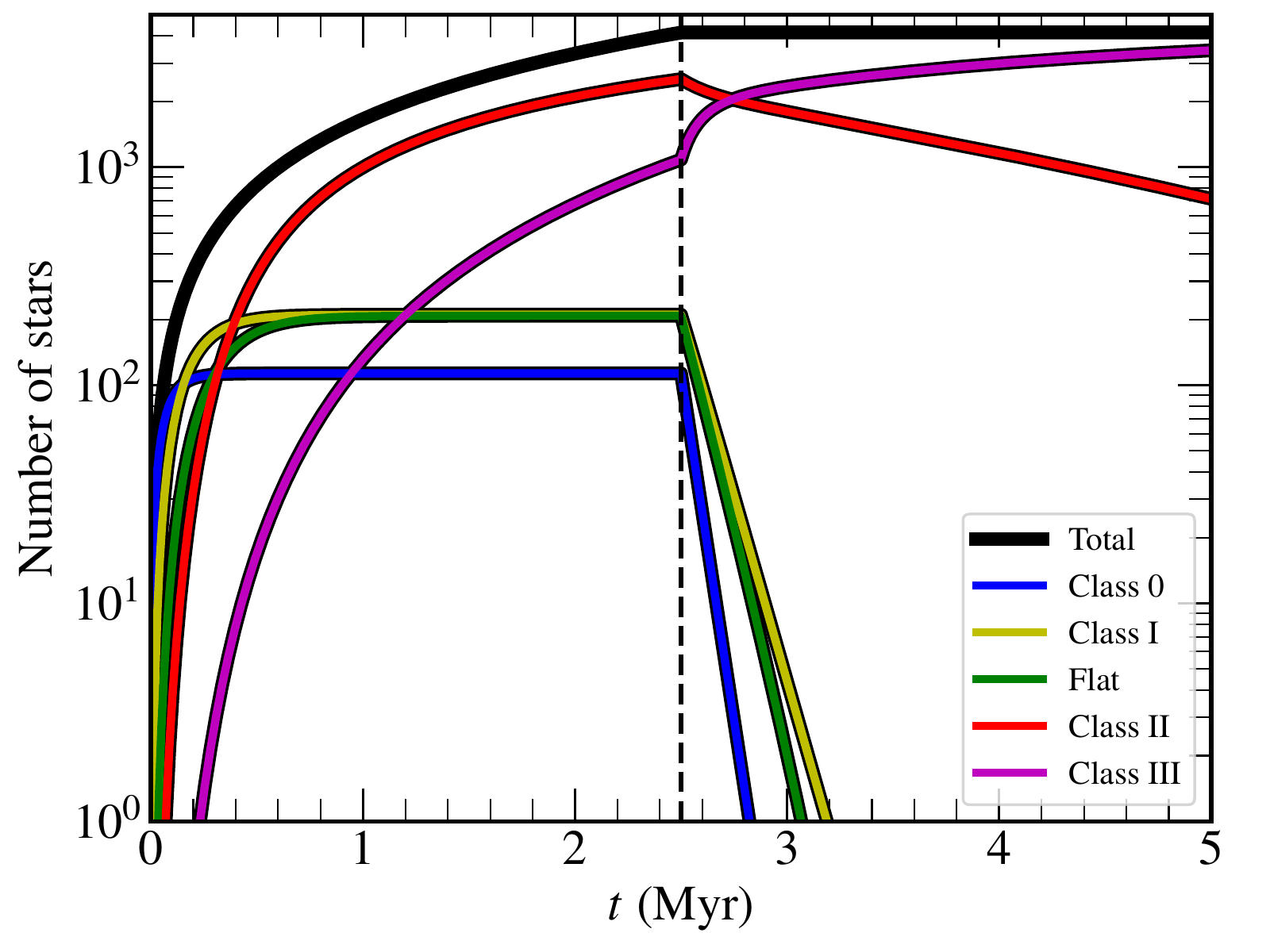}
\caption{\textit{Top:} Mean fraction of protostars in a given Class as a function of time. The thick semi-transparent lines in the background show the results for each cloud, whereas the black/colored lines show the results for the total population. Observed fractions are shown as dots. The fraction of Class III sources is obtained as $N_{\rm III}$ = $N_{\rm tot}$ -- $N_{\rm II}$ -- $N_{\rm flat}$ -- $N_{\rm I}$ -- $N_{\rm 0}$. \textit{Middle:}  Absolute number of stars in each Class as a function of time for all clouds together, for the star formation rate provided in Table \ref{tab:age}. \textit{Bottom:} Same as middle panel, but with a cut-off time of 2.5 Myr after which the star-formation rate is set to zero, as indicated by the vertical dashed line. When the rate is zero, the total number of stars naturally remains constant. 
\label{fig:evolution}}
\end{figure}

The resulting decay rates may be used for inferring the relative number of protostars at a given evolutionary stage as a function of time. Coupled with a star-formation rate, the decay rates can furthermore be used to provide the total number of protostars at a given stage. This evolution as a function of time is illustrated in Fig. \ref{fig:evolution}, where also the observed fractions are displayed. The absolute number of stars in each class in all clouds as a function of time is also displayed in Fig. \ref{fig:evolution}. The number of Class 0, I, and Flat sources is constant after a few half-lives ($<$ 0.5 Myr) because the half-life is so short that the decay to Class II sources proceeds at the same rate as Class 0 sources is injected into the cloud. Furthermore, there are about twice as many Class I and Flat sources as Class 0 sources, independent of time after $\sim$ 0.5 Myr. The constraining factor in determining the star-formation time of a cloud is therefore the number of Class II and III stars relative to Class 0, I, and Flat sources. Once the gas reservoir is spent and no more stars are forming, the Class 0, I, and Flat sources will rapidly decline, and only Class II and III sources will remain. This scenario is illustrated in Fig. \ref{fig:evolution}, where  no new Class 0 objects form after 2.5 Myr. At $t$ $\sim$ 3 Myr there are practically no Class 0, I, and Flat sources left in the cloud.

\section{Discussion: Protostellar half-lives}
\label{sec:disc}

The inferred decay rates and Class half-lives depend on a number of assumptions as listed in Sect. 2. Each of these main assumptions will be discussed below. Furthermore, the protostellar half-life or decay will be discussed as a concept, i.e., the underlying idea of an apparently stochastic process governing protostellar evolution rather than a deterministic process. Finally, the implications for estimating protostellar half-lives in general will be discussed. 

\subsection{Validity of assumptions}

Section 2 lists the main assumptions under which this work was performed. In the following, each of these discussions will be discussed in more detail. 

\subsubsection{Star formation is continuous}

The assumption that star formation is continuous and only determined by a single parameter, the star-formation rate (or $\lambda_{\rm SF}$) is difficult to validate directly from an observational point. While it is clear that clouds generally harbor sub-regions of increased density and star-formation activity (e.g., the young NGC1333 region in Perseus), it is unclear what the lifetime is of such sub-regions. It is possible that these represent regions of enhanced star-formation activity, and thus that the star-formation rate locally is higher and not representative of the entire cloud. 

On the other hand, without a complete picture of the evolution of a cloud, it is difficult to estimate how many previous epochs there have been of enhanced star-formation activity. Thus, any cloud may contain regions where the age of the cloud appears older (i.e., more Class II/III objects) or younger (more Class 0 objects), whereas the age that is inferred here is an average of the entire cloud. If a cloud is divided into sub-regions and these are analyzed separately, the number of protostars quickly decreases and the uncertainties on the decay rates increase dramatically. The only way to overcome this problem is to look at more massive clusters that are close enough that individual protostars can be observed directly. \citet{fischer17} examined the Orion Clouds, and concluded that there is indeed variation between different sub-regions. However, these authors also concluded that when averaged over the entire cloud, the star-formation rate appears to have been stable for the past 0.5 Myrs. Therefore the assumption of a constant star-formation rate appears reasonable.

\subsubsection{Only time matters for evolution}

One of the natural time scales for star formation is the free-fall time, which only depends on the enclosed mass and is typically 0.1 Myr for a 0.5 $M_\odot$ star \citep{myers98, young05}. This time scale therefore suggests that protostellar lifetimes depend on mass. On the other hand, it is well established that gravity is not the only force working on forming stars: other forces work to keep gravity at bay (magnetic fields, turbulence, thermal pressure, etc.). As with the first assumption, it is difficult to verify the validity of this assumption without breaking the protostellar sample into separate mass bins and thus increasing the spread or uncertainty on any inferred parameter. Moreover, protostellar masses are notoriously difficult to measure at the earliest embedded stages, where the only direct measurements come from observations of disks around such objects \citep[e.g.,][]{tobin12}. However, the sample of known Class 0 disks is rather limited, and examples of Class 0 sources without detected disks exist \citep[e.g.,][]{evans15}. Thus, breaking the sample up based on mass is currently not possible. 

The range of inferred star-forming times is very small, a factor of two between the youngest (Perseus) and oldest (Serpens) cloud. This small spread suggests two things. First, there is a natural bias in how the clouds have been selected: the clouds all contain many protostars ($>$ 100), they all harbor young Class 0 protostars, and they all form part of the Gould Belt. The location and age selection suggest that star formation in these parts of the Gould Belt has only been occurring for a few Myr at most. Second, there is very little difference between the clouds. The small differences also appear in the decay rates which, within the uncertainties, are the same for all clouds. Therefore the clouds analyzed here can be treated as a single cloud, which increases the accuracy of the inferred half-lives and decay rates. 

The number of Class 0 objects in the Ophiuchus cloud is very low \citep{enoch09}, and the assumption of 35\% Class 0 and 65\% Class I sources in the combined Class 0+I sample may be an oversimplification. If Eq.s 6--9 are solved for Ophiuchus with the observed Class 0 and I fractions \citep{enoch09}, the cloud age increases from 1.2 to to 1.3 Myr, and the Class 0, I, and Flat half-lives change to 10 kyr (30 kyr), 81 kyr (55 kyr), and 140 kyr (130 kyr), respectively, for a Class II half-life of 2 Myr and where the numbers in brackets are for the solution with 35\% Class 0 and 65\% Class I sources. While the Class I and Flat half-lives are within a factor of two of the average values, the Class 0 half-live is a factor of three lower. Moreover, the Class 0 half-life inferred for this cloud is already low, 30 kyr versus 50 kyr inferred for the combined sample. This seemingly very low number of Class 0 sources in Ophiuchus suggests, as has also been noted previously in the literature \citep{visser02, enoch09}, that this cloud is currently forming Class 0 objects at a very low rate. It is possible that the cloud is already at the turnover point where the star-formation rate will continue to decrease or star formation comes to a halt, in which case Eq.s 6--9 no longer apply. Instead, another free parameter enters the equations, and that is the cut-off time. Without further observational constraints, it is not possible to determine this parameter at present.

\subsubsection{Protostellar evolution is sequential}

This assumption is necessary for this particular method, as opposed to the number-counting method where, in principle, the different protostellar classes are treated independently. This assumption is valid as long as the observationally defined Class system corresponds to a physical evolution, i.e., the evolution from Class 0 to III corresponds to a similar evolution from a physical Stage 0 to III. Observed classifications are known to be erroneous sometimes \citep{dunham14}, for example if a Class I disk is observed edge-on, the object may seem redder than it actually is \citep{fischer17}. Different methods exist for verifying the classification of an object, and although these are typically observationally expensive \citep{vankempen09c}, \citet{heiderman15} observed more than 500 protostellar sources to classify them based on their molecular emission. They found that the Flat sources do not correspond to a distinct physical class, but rather these sources are divided roughly equally between Class 0+I and Class II sources. If that is indeed the case, the Class 0 and I half-lives change to 54 kyr (47 kyr) and 99 kyr (88 kyr). Thus, while the half-lives do increase, the change is almost negligible. Thus, the assumption may be changed to any source-to-source variations are washed out by the large number of protostars, and that, as a sample, the protostars are well behaved. 

Moreover, protostars may move back and forth between different observational Classes, particularly if the sources undergo episodic accretion \citep{dunham10}. The lifetimes that are inferred here are of the observed Classes, and it is clear that these do not necessarily translate to physical Stages. To obtain the link with the lifetime of protostars in a physical Stage, further and more complex modeling is required; such a modeling will need to incorporate a non-steady-state approach to obtaining lifetimes.

\subsubsection{Star-formation rate and half-lives are constant}

In this work, Class 0 objects are assumed to be injected or formed at a constant, but unknown, rate. This assumption is often made in similar studies \citep[e.g.,][]{fletcher94a, fletcher94b, myers12}, mainly because it is the simplest assumption which still provides useful estimates of half-lives \citep{myers12}. 

The assumption that the half-lives are constant, i.e., that the probability of decay is constant as a function of time, is motivated by two factors. First, we ideally want a parametrized distribution depending on a single parameter for simplicity. Second, we assume that the evolution of one protostar does not affect any other protostars, and that all protostars are born equal, i.e., only time matters for the evolution. Finally, any distribution must be able to account for the presence of evolved Class III objects in young clusters, e.g., Perseus with an estimated star-forming time of $\sim$ 1 Myr. The Poisson distribution naturally fits these motivating factors, and even though the data do not directly constrain the actual distribution, it is a good starting point. 

Any continuous smooth probability distribution will have a non-zero probability of decay at both $t$ = 0 and $\infty$. In practice, and certainly for clusters with $\sim$ 10$^3$ members, the actual number of protostars in the Class III stage will be vanishingly low at very early times. Similarly, the number of Class 0 protostars surviving for many millions of years will be negligible. Thus, although the probability of decay is the same at all times, in practice the number of protostars in each Class is well-behaved. 

Instead of a constant decay probability, the other extreme may be examined, where all sources in a given Class have a zero probability of decay until a fixed time $t_0$ where the probability jumps to 1. In such a model, Class 0 protostars form at the constant star-formation rate until the Class 0 lifetime is reached. At this time, Class I protostars begin forming at the star-formation rate because the Class 0 protostars that formed at $t$ = 0 now decay to Class I. This also means that the number of Class 0 protostars becomes constant, because the Class 0 formation rate is balanced by the decay rate. The same results naturally apply to the other decays as well. In this model, the inferred protostellar lifetimes revert to the steady-state values, but the star-forming time of the clouds becomes the sum of the lifetime of each Class plus the time over which Class III sources have formed. For the Gould Belt clouds, that number is 3.2 Myr for an assumed Class II lifetime of 2 Myr. Two important conclusions are reached in this particular model. First, Class I sources only start forming after the Class 0 lifetime, and Flat sources only begin to form after the Class 0 and Class I lifetimes, etc. Thus, in the first 2.5 Myr of the cloud's star-forming time, there are no Class III sources in the cloud. Second, the star-forming time of the clouds is significantly longer than otherwise inferred, with correspondingly lower star-formation rates. 

The two models presented in this subsection, constant probability as a function of time or delta function at a given time, clearly represent two extremes. However, they serve an illustrative purpose: in the former model, the protostellar lifetimes are shorter than in steady-state but the cloud star-forming is consistent with other estimates; in the latter, the protostellar lifetimes line up with the steady-state solutions, but at the cost of star-forming time. Other probability distributions will likely provide values of the lifetimes and star-forming times between those reported here. If that is the case, then it is likely that the actual protostellar lifetimes are shorter than currently inferred in the steady-state solution, and star-forming times are longer than inferred.

\subsubsection{Populations are complete}

There is naturally some cut-off below which the sample of protostars is incomplete. This cut-off appears in two ways: either the stars are too weak to be detected, or they overlap at the resolution of \textit{Spitzer}. The sensitivity cut-off is particularly important for very low-luminosity objects. At the \textit{Spitzer} wavelengths, these appear as either very low-mass objects such as proto-brown dwarves, or as highly reddened objects with very little continuum emission below wavelengths of $\sim$ 100 $\mu$m. Such objects are similar to the recently described PACS Bright Red Sources \citep[PBRSs,][]{stutz13}. Based on number statistics, \citet{tobin16} estimate that Perseus should contain only 1--2 objects of a total of 377 protostars, and we therefore estimate that the number of PBRSs in the dense Gould Belt clouds is negligible ($\ll$ 1 \%).

\subsubsection{The final stable state is Class III}

Main sequence stars are the stable end product of protostellar evolution, but it is unclear how many main sequence stars there are in any cloud. The reason for this lack of knowledge is straightforward: each cloud is surrounded by a large number of stars, and determining if a star belongs to a cloud, is a foreground or a background star is challenging. The GAIA mission may provide answers to this question. By measuring the 3D velocities of all stars in the direction of a cloud, GAIA will estimate if a star is at the same distance as the cloud, and is comoving with the cloud. Spectroscopy may further be used to verify if a star is young or not \citep[e.g.,][]{rigliaco16}. If that is the case, there is a high likelihood that the star is a main-sequence star that formed in the cloud. 

If the number of main-sequence stars in a cloud, or the entire Gould Belt, is known, $N_{\rm III}$ can be solved for in Eq.s 6--9, and the decay rate of Class III sources can be inferred. As discussed by \citet{dunham15}, the combined half-life for the Class II+III stages may be a better measure for protostellar half-lives than just the Class II stage (see below), and, with a knowledge of the main sequence population, the half-lives can be scaled to this number rather than the Class II half-life. However, until the main sequence population is known, the half-life approach to protostellar lifetimes is only possible if there is a final stable state and that is, by necessity, the Class III stage for the moment.

\subsubsection{The Class II half-life is 2 Myr}

\begin{figure}
\center\includegraphics[width=0.9\columnwidth]{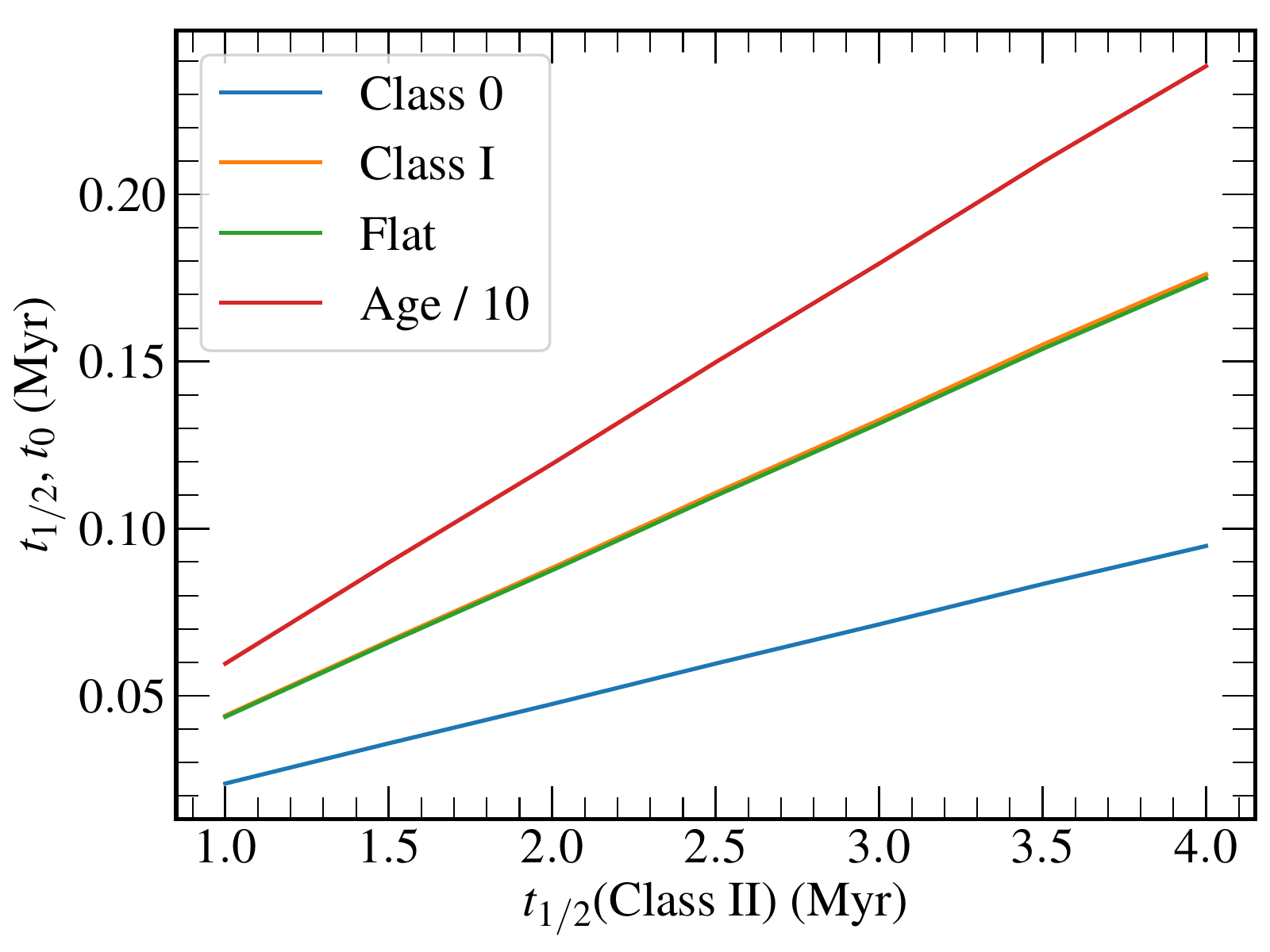}
\caption{The change in Class 0, I, and Flat half-lives and cloud age, $t_0$, as a function of $t_{1/2}$(Class II), ranging from the most extreme values in either end, 1 and 4 Myr. 
\label{fig:lambda_var}}
\end{figure}

\citet{dunham15} argue that the currently best estimate for protostellar half-lives is the combined Class II and III half-life which is 3 Myr. As argued above, the combined Class II+III half-life is inconvenient for the half-life method to work, but instead the Class II half-life can be assumed to be 3 Myr instead of 2 Myr. The other half-lives scale linearly with the assumed Class II half-life, as illustrated in Fig. \ref{fig:lambda_var} where the Class 0, I, and Flat half-lives are shown as a function of the Class II half-life. If the Class II half-live increases by 50\% to 3 Myr, all other half-lives increase by 50\% as well. The key point here is not so much the actual value of the Class II half-life, but rather that the ratio of the Class 0, I, and Flat half-lives to the Class II half-life is (2.4$\pm$0.2)\%, (4.4$\pm$0.3)\%, and (4.3$\pm$0.4)\%, respectively, and that these are constant under the assumption of constant star-formation rate. The other main implication is that protostars spend $\sim$ 10\% of their time on average in the embedded phase (Class 0, I, Flat stages) and the remaining 90\% of their time as Class II or III objects. 

The main reason for adopting a Class II half-life of 2 Myr here is that the half-lives are easily compared to other studies. Most studies find Class 0 lifetimes exceeding 0.1 Myr, i.e., a factor of at least two greater than the half-life reported here \citep{kenyon90, enoch09, dunham15}, and a total duration of the embedded phase (Class 0+I) of $\sim$ 0.5 Myr \citep{wilking89, kenyon90, hatchell07, evans09, enoch09, dunham15}, or three times longer than reported here. The main reason is that the number of Class II sources is typically the highest in any cloud, and these Class II sources evolved from Class 0, Class I, and Flat sources, i.e., any Class II source will have to have passed through these stages first. Class 0 objects are injected into the cloud at a constant rate, and these Class 0 objects need to be removed fast enough that there is not a build-up. Furthermore, enough stars need to have moved through these different stages to end up at the Class II stage in sufficient numbers at the given age of the cloud. This sequential evolution is not explicitly considered when using the number counts to infer ages, where the implicit assumption is that the relative populations are in a steady-state solution. If the half-life of any single stage is significantly shorter than the cloud lifetime, that assumption is valid. Such a scenario does not apply to the young, dense clouds in the Gould Belt, and a sequential analysis is required as the one used here. 

It is worth noting that as time goes to infinity, the number of protostars in each class apart from the stable class becomes constant (see Fig. \ref{fig:evolution} for the cases of Class 0, I, and Flat sources). If the protostellar lifetimes are inferred in this regime based on the number-counting method, the same values for the lifetimes will be obtained.

\subsection{Half-life as a concept}

Adopting a half-life approach to protostellar lifetimes introduces a stochasticity to these lifetimes as opposed to protostellar evolution being deterministic. It is not given that a protostar will be in the Class 0 phase for $X$ number of years before moving on to the Class I phase for $Y$ years. Instead, the average protostar will spend $\sim$ 50,000 years as a Class 0 object, with a spread proportional to the square root of the half-life, and similarly for the other stages. Note that the uncertainty given in Table \ref{tab:decay} is not the spread in protostellar half-lives, but rather the uncertainty on the actual half-life. 

While the formalism for sequential nuclear decay relies on all nuclei being born equal, and that only time determines the likelihood of a decay, protostars are not born equal. They form with different masses, in more or less turbulent media, exposed to different UV or magnetic fields which affect the temperature structure, their density structures are different, etc. Furthermore, the evolution of one protostar may affect the evolution of others, e.g., through UV output. All these factors influence the protostellar evolution to some degree and will essentially mean that no two protostars are perfectly identical. Moreover, protostars undergo episodic accretion, and they are observed at different inclination angles; these two effects may affect how objects are classified \citep[e.g.,][]{dunham15, fischer17}. However, it is also clear that when considering the protostars as a sample, these effects will make protostellar evolution appear random: even with the best physical models we are only now beginning to fully understand the protostellar collapse and subsequent evolution in detail \citep{kuffmeier16}. Thus, while it is clear the different physical conditions add to the spread in actual protostellar lifetimes, this spread appears stochastic unless all the underlying processes and the history of the cloud material is accounted for, which is practically impossible.

\subsection{Implications for star-formation rates}

The star-forming times of the individual clouds are mainly constrained by the number of Class II and III objects, the two long-lived stages (2 Myr and pseudo-stable, in this study). This implies that the inferred cloud lifetimes are not that different from what has been inferred elsewhere in the literature based on disk counting \citep{mamajek09}. This implication also means that the star-formation rates inferred here are consistent with other estimates. As an example, \citet{mercimek17} report a star-formation rate for Perseus of $\sim$ 150 $M_\odot$ Myr$^{-1}$ vs. 190$\pm$40 $M_\odot$ Myr$^{-1}$ found here (Table \ref{tab:age}). Their inferred star-formation rate is based on observations of primarily prestellar cores and Class 0+I objects and is thus independent of the method used here. \citet{shimajiri17} uses a similar method to the one used here, and assumes a mean stellar mass and cloud age of 2 Myr to find SFRs that are within a factor of 2--3 those reported here. The major difference is the different cloud ages used, and the number of protostars considered in each sample. However, the key part here is that not only does the half-life method presented here provide half-life estimates of each Class, but also a total duration of the star-formation time in each cloud.

\section{Summary and conclusions}

We have presented a new method for inferring half-lives for each protostellar Class, based on the formalism of sequential nuclear decay with a constant injection rate. This method naturally provides a distribution, rather than a single lifetime of each Class, and naturally provides uncertainties and spreads on the half-lives. Moreover, the method provides a means for estimating the number of protostars in each Class as a function of time. The method only relies on very basic assumptions, which are not too dissimilar to the assumptions made when counting protostars in each Class and comparing them to the assumed lifetime of a single Class. The main difference between the methods is whether protostellar populations are assumed to be in a steady-state or non-steady-state solution. Although the specific distribution cannot be directly inferred from observations, the new method is an important step toward looking at protostellar populations, and related time-scales, out of steady state.

The half-lives scale linearly with the Class II half-life, and the key numbers are that the fraction of the half-lives is (2.4$\pm$0.2)\%, (4.4$\pm$0.3)\%, and (4.3$\pm$0.4)\% for Class 0, I, and Flat sources with respect to the Class II half-life. For a Class II half-life of 2 Myr, these fractions correspond to 50, 90, and 90 kyr. The half-lives are significantly shorter than the lifetimes estimated from the number-counting, steady-state method \citep[typically 150, 300, 300 kyr, respectively,][]{dunham15}; the main reason is that the half-life needs to be short enough for a significant fraction of the Class II protostars to have gone through the initial Class 0, I, and Flat stages. If the probability is assumed to be a delta function instead of constant, the steady-state values for protostellar lifetimes are recovered, but at the cost of star-forming time; the clouds will have to have formed stars for significantly longer than is presently inferred. We emphasize that the values inferred here depend critically on the assumed Class II half-life, and on the assumption that the star-formation rate is constant over time. This comparison highlights the importance of looking at the star-formation process as sequential and not as steady state. 

\begin{acknowledgements}
We would like to thank the referee, Prof. J. Hatchell, and Prof. N.J. Evans II for their very constructive comments which served to improve this manuscript. Sub-millimeter astronomy in Copenhagen is supported by the European Research Council (ERC) under the European Union's Horizon 2020 research and innovation programme (grant agreement No 646908) through ERC Consolidator Grant ``S4F''. Research at the Centre for Star and Planet Formation is funded by the Danish National Research Foundation. 
\end{acknowledgements}

\bibliographystyle{aa}
\bibliography{bibliography}

\end{document}